\documentclass[aps, prl, reprint, superscriptaddress, amsmath, amssymb]{revtex4-2}
\usepackage{epsfig}
\usepackage{epstopdf}
\usepackage{multirow}
\usepackage{graphicx}
\usepackage{dcolumn}
\usepackage{bm}
\usepackage{textcomp}
\usepackage{array}
\usepackage{csquotes}
\usepackage{subcaption}
\usepackage{booktabs}
\usepackage{amsmath}
\usepackage{hyperref}
\hypersetup{
    colorlinks=true,
    linkcolor=blue,
    filecolor=magenta,      
    urlcolor=red,
    citecolor=blue,
    pdftitle={},
    }
    
\usepackage{tikz}
\newcommand{\orcidicon}{%
    \begin{tikzpicture}
        \draw[lime, fill=lime] (0,0) circle [radius=0.16] node[white] {{\fontfamily{qag}\selectfont \tiny ID}};
        \draw[white, fill=white] (-0.0625,0.006) circle [radius=0.007];
    \end{tikzpicture}
    \hspace{-2.9mm}
}
\foreach \x in {A, B}{\expandafter\xdef\csname orcid\x\endcsname{\noexpand\href{https://orcid.org/\csname orcidauthor\x\endcsname}{\noexpand\orcidicon}}}

\newcommand{\iitmsps}{School of Physical Sciences, Indian Institute of Technology Mandi, Kamand - 175075, India}
\newcommand{\iitm}{School of Mechanical and Materials Engineering, Indian Institute of Technology Mandi, Kamand 175075, India}

\begin{document}
\preprint{APS/123-QED}
\title{Multiple topological electronic and phononic quasiparticle\\ excitations in hexagonal TTe (T=Hf, Zr \& Ti) crystals}

\author{Prakash Pandey \orcidA{}}
\email{prakashpandey6215@gmail.com}
\affiliation{\iitmsps}
\author{Sudhir K. Pandey \orcidB{}}
\email{sudhir@iitmandi.ac.in}
\affiliation{\iitm}


\date{\today}

\begin{abstract}

The exploration of topological fermions and bosons marks a new chapter in condensed matter physics, unveiling rich and unconventional phenomena. 
Particularly in quantum field theory, exotic quasiparticle excitations such as Dirac and Weyl fermions can serve as direct analogs, and their recent experimental realizations have sparked significant interest in these topological quasiparticles. Although there are various reports on the coexistence of unconventional fermionic quasiparticles such as spin-1/2 (type-I, type-II, \& type-III), spin-1 (threefold degeneracy), nodal line (type-I, type-II, \& type-III) and others, reports on the simultaneous presence of such quasiparticles in both electronic and phononic spectra within a single material remain extremely limited. 
Herein, using \textit{state-of-the-art} \textit{ab initio} calculations, we propose the HfTe class of materials, which hosts coexisting type-I, type-II, and type-III Weyl and nodal line phases, along with pseudospin-1/spin-1 quasiparticles in both electronic and phononic states.
We have found that these excitations are robust against variations in exchange-correlation functionals, spin-orbit coupling, and lattice parameters, confirming that multiple topological phases in this class of materials are likely to be observed experimentally.

\end{abstract}

\maketitle

{\it Introduction.|}
The novel topological phenomena in the electronic and phononic systems are currently one of the most emerging fields in condensed matter physics and material sciences~\cite{RevModPhys.90.015001, RevModPhys.93.025002}. 
Although the progress in understanding unconventional quasiparticles beyond Dirac~\cite{10.1126/science.1245085} and Weyl~\cite{10.1126/science.aaa9297} in conventional crystals is highly exciting, the coexistence of such quasiparticles in both electronic and phononic states of solid crystalline materials is of fundamental importance.
Besides the well-known Dirac ($\texttt{C}$=0) and Weyl fermions ($\texttt{C}$=$\pm$1), theorists have predicted other unconventional quasiparticles, including charge-2 fourfold ($\texttt{C}$=$\pm$2)~\cite{10.1126/science.aaf5037}, Rarita-Schwinger-Weyl ($\texttt{C}$=$\pm$4)~\cite{PhysRevLett.119.206402}, charge-4 sixfold/double three-component ($\texttt{C}$=$\pm$4)~\cite{10.1126/science.aaf5037}, and eightfold double Dirac fermions ($\texttt{C}$=0)~\cite{PhysRevLett.116.186402}. These exotic quasiparticle excitations are unlocking potential applications in next-generation energy and electronic devices~\cite{wang2020topological, liu2020topological}.
Parallel to topological classifications in electronic systems, several topological states of phonons, including Weyl, Dirac, nodal-ring, and higher-order phonons, have also been theoretically proposed and experimentally verified in real materials~\cite{PhysRevLett.95.155901, PhysRevLett.114.114301, PhysRevLett.117.068001, PhysRevLett.116.135503}.

Furthermore, topological phononic states exhibit novel applications in high-efficiency phononic circuits/diodes~\cite{PhysRevLett.93.184301, 10.34133/2019/5173580}, phonon scattering, electron-phonon interactions (EPIs)~\cite{HAN2023520, PANDEY2023415301}, and unconventional heat transfer. For example, topological phonons can also induce one-way edge phonon states, similar to topological edge states in electrons~\cite{PhysRevB.96.064106}. These states conduct phonons with negligible or zero scattering~\cite{he2016acoustic, fleury2016floquet}, enabling the design of phononic diodes~\cite{PhysRevLett.93.184301, 10.34133/2019/5173580}. 
Another promising example of topological phonons is their ability to enhance the thermoelectric properties of materials~\cite{PhysRevMaterials.2.114204}, as gapless topological phonon modes can increase phonon-phonon scattering channels, significantly decreasing the mean free path and suppressing phonon thermal conductivity~\cite{PhysRevB.100.245203, PhysRevMaterials.2.114204}.
In particular, the advantage of topological phonons can be appreciated due to various special features, including band touching and their associated excitations, which can be found anywhere in the entire frequency range of the phonon spectrum. Additionally, it is possible to probe topological states in both the acoustic and optical frequency regimes using experimental techniques. However, in fermionic states, this is mostly possible only in the region near the Fermi energy within band structures. Similar to electronic states, topological phonons also induce drumhead-like surface states~\cite{deng2019nodal} and anomalous transport behaviors, such as the anomalous phonon Hall effect~\cite{PhysRevLett.105.225901, PhysRevX.12.041031}.

The coexistence of multiple topological phases and their transitions in both electronic and phononic states within single quantum material represents one of the most active and fruitful areas of current research, due to its significance from both fundamental and application perspectives. 
However, the simultaneous presence of multiple topological electronic and phononic quasiparticle excitations such as spin-1/2 with type-I, type-II, \& type-III features; nodal lines with type-I, type-II, \& type-III features; and spin-1 quasiparticles, without any external perturbation, remains scarce despite extensive efforts.
A real material that hosts multiple topological phononic quasiparticle excitations at lower frequencies and with fewer atoms in the primitive unit cell would not only provide an opportunity to study exotic quasiparticle excitations but also offer a platform to explore the emergent physics arising from the interplay between topological excitations and EPIs.
For example, low-frequency modes may significantly contribute to EPIs and thereby affect thermal conductivity, revealing a strong connection between topological and thermophysical properties~\cite{PhysRevMaterials.2.114204}. Therefore, the search for simpler systems that exhibit multiple topological electronic and phononic quasiparticle excitations, along with relatively low phonon frequencies, is urgently needed to advance current technological applications.
In the present work, we explore the electronic and phononic properties of WC-type materials (HfTe, ZrTe, and TiTe) based on \textit{state-of-the-art ab-initio} calculations.
Based on our findings, we propose that the HfTe possesses multiple topological electronic and phononic quasiparticle excitations. Their robustness against variations in exchange-correlation functionals, the strength of spin-orbit coupling (SOC), and small changes in lattice parameters suggests that their existence can also be confirmed experimentally.

{\it Computational details.|}
The \textit{ab-initio} calculations are performed within the framework of density functional theory (DFT) using the augmented plane wave plus local orbitals (APW+lo) method, as implemented in the \texttt{WIEN2k} package \cite{blaha2020wien2k}. The generalized gradient approximation (GGA)~\cite{PhysRevLett.77.3865}, within the Perdew-Burke-Ernzerhof (PBE) formulation, is employed as the exchange-correlation (XC) functional for calculating the ground-state energy of TTe (T = Hf, Zr, Te). This XC functional is used to optimize the lattice constant of TTe, and the optimized lattice constant for each material is given in Table I of the Supplemental Material (SM)~\cite{spl}. 
A 16$\times$16$\times$16 $k$-mesh size is used for the self-consistent ground state energy calculations. Furthermore, the energy convergence threshold is set to $10^{-8}$ Ry/cell. \texttt{PY-Nodes}~\cite{PANDEY2023108570} is used to calculate topological features such as nodal points and nodal lines. This code is based on the Nelder-Mead function minimization method~\cite{10.1093/comjnl/7.4.308}. To calculate the topological charge or chirality for each Weyl point, we have used the \texttt{WloopPHI} code~\cite{saini2022wloopphi}.
Furthermore, to examine the robustness of the topological phase of HfTe with respect to the strength of SOC, we performed electronic structure calculations for different SOC strengths. Since the SOC Hamiltonian is inversely proportional to the square of the speed of light, decreasing the speed of light ($c$) could enhance the strength of SOC~\cite{blaha2020wien2k}. Therefore, $c$ is treated as a parameter, and its value is reduced in order to increase the strength of SOC in the calculations. The ground state energy calculations in the presence of SOC are performed using three different values of $c$: 137.03, 135.65  \& 134.28 a.u.
The detailed procedure for assessing the robustness of the topological phase against variations in SOC strength was performed in our previous works~\cite{pandey2025realization, Pandey_2025_JPD}.
Apart from this, to examine the robustness of the topological phase of HfTe against chemical doping, we performed electronic structure calculations for different doping combinations, including HfTe$_{0.875}$Se$_{0.125}$, HfTe$_{0.75}$Se$_{0.25}$, HfTe$_{0.625}$Se$_{0.375}$, and HfTe$_{0.5}$Se$_{0.5}$. Supercells of HfTe$_{1-x}$Se$_x$ containing 16 atoms are constructed. After constructing HfTe$_{1-x}$Se$_x$ ($x = 0.125$, $0.250$, $0.375$, $0.5$), atomic relaxation and structural optimization are performed to minimize the forces. The equilibrium lattice parameters are obtained by minimizing the total energy of the crystal, calculated for different values of the lattice constant.

As the counterpart of the electron, the phonon is quantized excited vibrational states of interacting atoms. It is one of the most common quasiparticles that plays an important role in understanding the phononic properties of the material. The phonon dispersions of TTe are calculated using APW+lo method with the \texttt{WIEN2k} package and the \texttt{PHONOPY} package~\cite{togo}. A $2 \times 2 \times 2$ supercell size is adopted, together with an $8 \times 8 \times 7$ \textbf{k}-mesh size. Furthermore, the force convergence limit for the self-consistent method is set to $10^{-6}$ Ry/Bohr. To calculate nodal line/point phonons, we have used the \texttt{PH-NODE} code~\cite{PANDEY2024109281}. \\\\\\\\\\\\\\\\

\begin{figure}[h!]
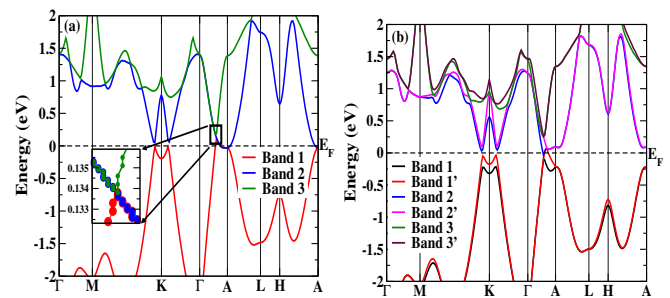
 
\centering
\includegraphics[width=0.49\linewidth, height=4cm]{Figure_eps/HfTe.bands_no_soc.eps}
\includegraphics[width=0.49\linewidth, height=3.7cm]{Figure_eps/HfTe.bands_soc.eps}
\caption{\label{Fig.band}\footnotesize{(Color online) The bulk band structure of HfTe along high-symmetric points in the first BZ (a) in the absence of SOC (b) in the presence of SOC.}}
\end{figure}

\begin{figure}[h!] 
\centering
\includegraphics[width=0.32\linewidth, height=3.5cm]{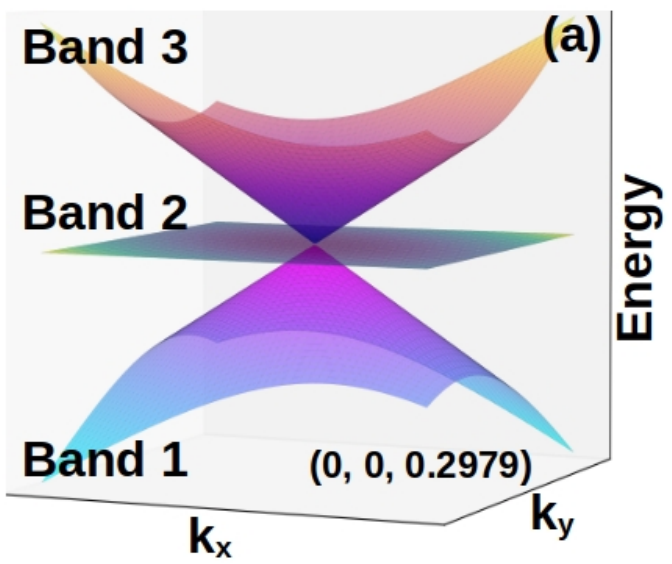}
\includegraphics[width=0.32\linewidth, height=4.0cm]{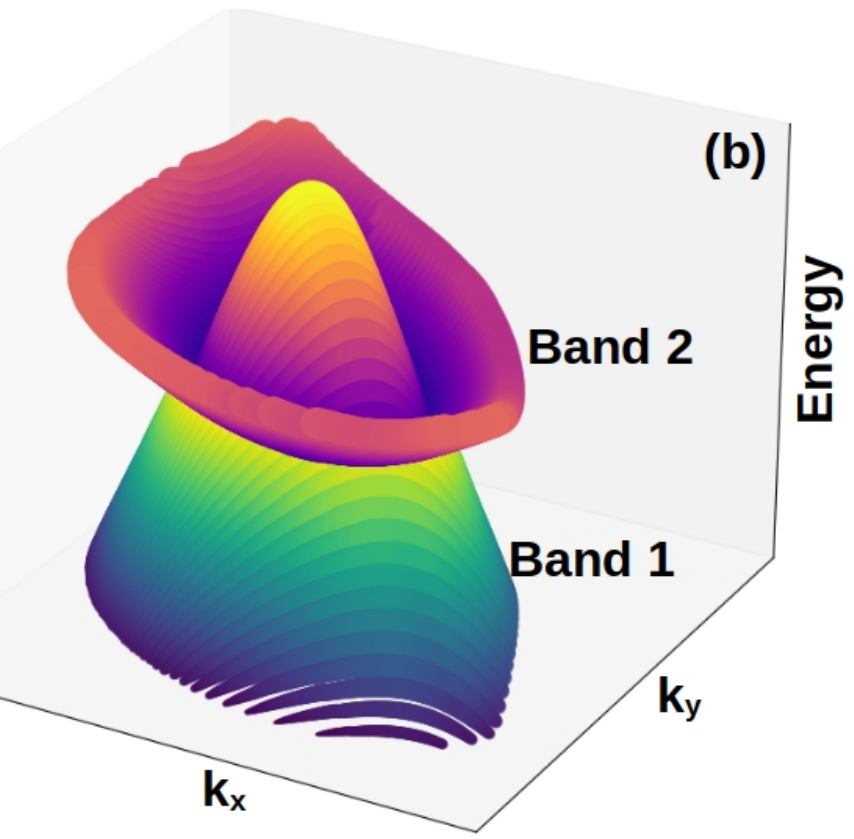}
\includegraphics[width=0.32\linewidth, height=3.5cm]{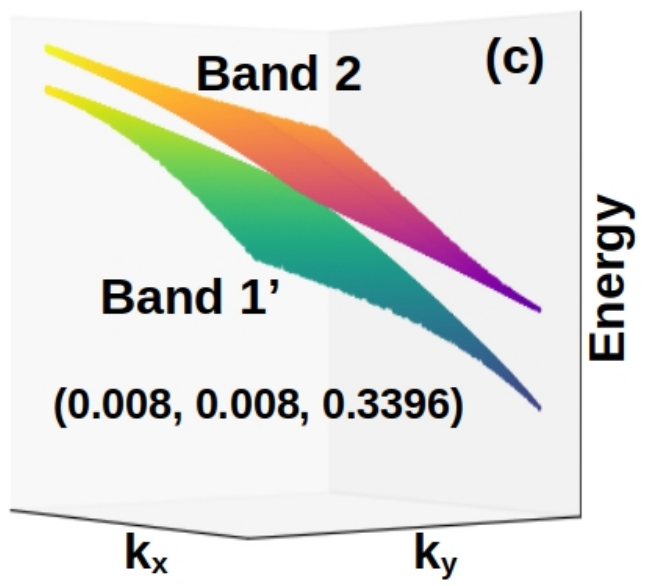}
\includegraphics[width=0.32\linewidth, height=3.5cm]{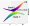}
\includegraphics[width=0.32\linewidth, height=3.5cm]{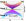}
\includegraphics[width=0.32\linewidth, height=3.5cm]{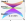}
\includegraphics[width=0.48\linewidth, height=3.5cm]{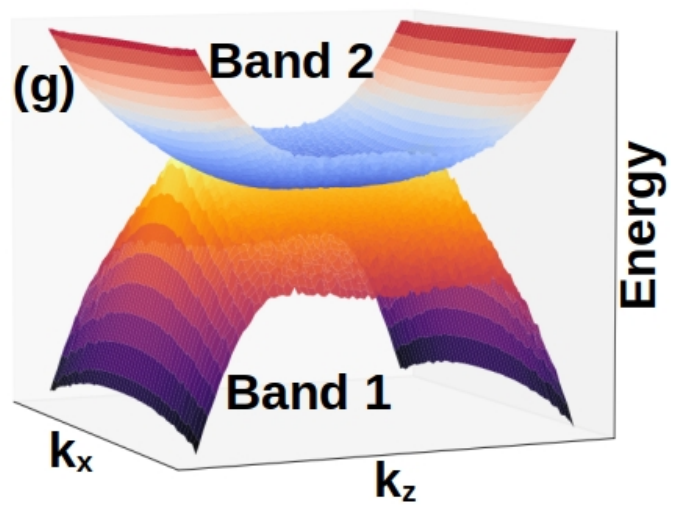}
\includegraphics[width=0.48\linewidth, height=3.5cm]{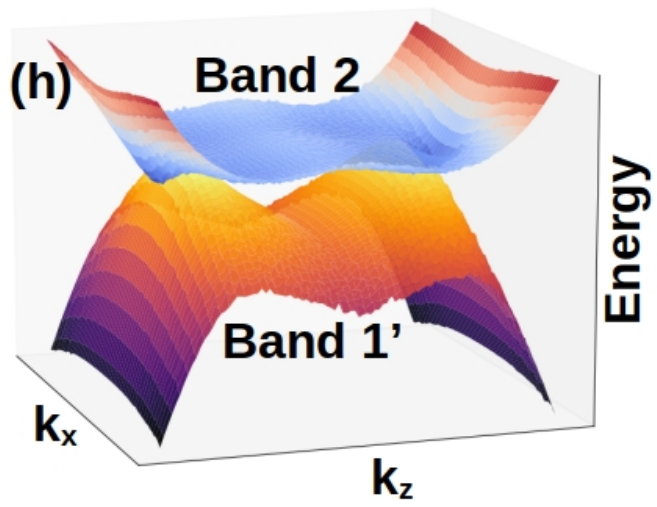}
\caption{\label{Fig.soc_type}\footnotesize{(Color online) 3D representation of the energy dispersions of HfTe around the wave vector \textbf{k} (a)-(b) pseudospin-1 threefold degenerate fermions and type-I nodal line fermions in the $k_x$-$k_y$ ($k_z$=0) plane when SOC is excluded (c)-(d) type-II and type-I Weyl fermions in the presence of SOC (e) ((f)) pseudospin-1 threefold degenerate fermions between bands 1, 1', and 2 (1', 2, and 2') when SOC is included (g) ((h)) symmetry-enforced straight nodal line along $\Gamma$-A high-symmetric path in the absence (presence) of SOC.}}
\end{figure}

Since single crystals of ZrTe have already been synthesized~\cite{orlygsson1999crystal, de1997phase} and possess hexagonal crystal symmetry with space group \texttt{P$\bar{6}$m2} (No. 187), the other two materials, HfTe and TiTe, may also be prepared and crystallize in the same space group.
Recently, the WC class of materials has been theoretically reported to host topological excitations in electronic (phononic states), including nodal lines, triply degenerate nodal points, and Weyl fermions (type-I Weyl and triply degenerate nodal point phonons)~\cite{PhysRevB.99.174306, PhysRevB.97.054305}. In this class, the materials belong to space group \texttt{P$\bar{6}$m2} (No. 187), featuring two inequivalent Wyckoff sites 1a (0, 0, 0) and 1d (1/3, 2/3, 1/2) in a noncentrosymmetric hexagonal structure. The space group \texttt{P$\bar{6}$m2} lacks inversion symmetry and possesses only time-reversal symmetry. Therefore, the absence of inversion symmetry is expected to give rise to a Weyl nodal line in the absence of SOC.  

{\it Topological fermions without and with SOC.|}
We have performed the \textit{ab-initio} calculations of HfTe within generalized gradient approximation. 
Figure \ref{Fig.band}(a) and (b) illustrate the calculated electronic band structure of the bulk system without and with SOC, respectively.
From Fig. \ref{Fig.band}(a), it is seen that around the Fermi level, bands 1, 2, and 3 are degenerate along the $\Gamma$-A path. This degeneracy of bands is further examined using \texttt{PY-Nodes}~\cite{PANDEY2023108570} code to confirm its authenticity. 
Since the \texttt{PY-Nodes} code is based directly on first-principles DFT methods, it is better equipped to identify all the nodal points present in the material~\cite{pandey2025revisiting}.
From the code, we have found a threefold degenerate fermion (TDF) at (0, 0, 0.2979), located 133.75 meV above the Fermi level ($E_F$). 
To explore its nature, we have plotted the 3D energy spectrum with respect to the wave vector \textbf{k} around the TDF (see Fig. \ref{Fig.soc_type}(a)). Such a TDF with one flat energy band in a fermionic system is commonly referred to as a pseudospin-1 fermion, in which the topological charges of bands 1, 2, and 3 are -2, 0, and 2, respectively.
Upon further investigation into the existence of a topological phase between bands 1 and 2, we performed calculations and found a nodal line phase featuring six nodal loops in the $k_z$=0 plane (see details in the SM~\cite{spl}). We have also examined the nature of the obtained nodal line and found two bands with opposite slopes in the directions normal to the ring, indicating a type-I nodal line (see Fig. \ref{Fig.soc_type}(b)). Since HfTe lacks inversion symmetry and possesses only time-reversal symmetry, it becomes Weyl nodal line semimetal in the absence of SOC.
The SOC term splits band 1 into 1 and 1', band 2 into 2 and 2', and band 3 into 3 and 3' along the $\Gamma$-M-K-$\Gamma$-A-L-H high-symmetric path, as shown in Fig. \ref{Fig.band}(b).
With the inclusion of SOC, each nodal line splits into four node points, resulting in a total of 24 node points from six nodal lines (see Table II in the SM~\cite{spl} for their coordinates and corresponding energies). 
Among the 24 node points, 12 are located 40.70 meV below $E_F$, and the remaining 12 are 58.47 meV above $E_F$. 
The coordinates of the node points at 40.70 meV and 58.47 meV are \(\left(\frac{2\pi}{a}(0.2675), \frac{2\pi}{b}(0.2675), \frac{2\pi}{c}(0.0246)\right)\) and \(\left(\frac{2\pi}{a}(0.0080), \frac{2\pi}{b}(0.0080), \frac{2\pi}{c}(0.3396)\right)\), respectively. 
Note that each node points has six symmetry-related pairs. To determine whether these node points are Weyl points, we calculated their topological charges ($C$) and found that $C \in \left\lbrace -1, 1\right\rbrace$. 
In order to determine the nature of each Weyl point, we have calculated the 3D energy spectrum around each node individually. The corresponding results are shown in Figs. \ref{Fig.soc_type}(c) and (d). Such band dispersions around the Weyl point (0.0080, 0.0080, 0.3396), as shown in Fig. \ref{Fig.soc_type}(c), are commonly referred to as type-II Weyl points. However, near the Weyl point (0.2675, 0.2675, 0.0246), the 3D energy spectrum exhibits type-I Weyl fermions (see Fig. \ref{Fig.soc_type}(d)). At the same time, we found that HfTe hosts two pseudospin-1 fermions: one formed by bands 1, 1', and 2 , and another by bands 1', 2, and 2'.
These two pseudospin-1 fermions are located at \(\left(0, 0, \frac{2\pi}{c}(0.2875)\right)\), with energy 56.67 meV below $E_F$, and at \(\left(0, 0, \frac{2\pi}{c}(0.3447)\right)\) , with energy 61.93 meV above $E_F$ (see Figs. \ref{Fig.soc_type}(e) and (f)). Such pseudospin-1 fermionic excitations carry Chern numbers of $\pm 2$; thus, some exotic physical phenomena, such as the bulk photogalvanic effect, may be observed. 
Threefold spin-1 fermion quasiparticles are expected to produce a quasi-linear energy dependence in the optical conductivity spectrum. For example, a recent study~\cite{10.1073/pnas.2010752117} reports that threefold spin-1 fermions exhibit an approximately linear frequency dependence of the interband optical conductivity at low energies, which is regarded as a hallmark signature of spin-1 fermions. Specifically, in CoSi~\cite{10.1073/pnas.2010752117}, the presence of a spin-1 fermion at the $\Gamma$ point gives rise to a kink in the optical conductivity around 0.2~eV, separating contributions from different topological quasiparticles and providing a distinct spectroscopic signature.
Interestingly, in addition to the above topological excitations, we also find straight nodal-line fermions along the $\Gamma$-A high-symmetric path in both the absence and presence of SOC (see Figs. \ref{Fig.soc_type}(g) and (h)). These straight nodal lines are symmetry-enforced, remaining robust even in the presence of SOC, and have been studied in various reports~\cite{PhysRevB.104.024304, PhysRevB.101.024301}.

To unveil the exotic physics of topological phases present in the electronic states of HfTe, it is also important to examine the robustness of the identified multiple topological phases against variations in lattice parameters, exchange-correlation (XC) functionals, SOC strength, and chemical doping. In this line, we first study the robustness of the topological phases against variations in the lattice parameters by $\pm$1\% and found that the topological excitations remain robust within this range (See Figs. 2 and 3 in the Supplementary Material (SM)~\cite{spl} for details about the band structure at $\pm$1\% without and with SOC). Next, we performed electronic structure calculations using the TB-mBJ potential, as it typically provides a band gap that is slightly closer to the experimental value~\cite{PhysRevB.99.035139}. Again, we found the above-discussed topological phases to be robust against different exchange-correlation functionals (for details, see Table III and Fig. 3 in the SM~\cite{spl}). 
Now we focus on the robustness of the topological excitation against variations in the strength of SOC, as DFT calculations are seen to slightly underestimate SOC effects compared to experimental results~\cite{PhysRevB.100.075205, DASILVA2021138896}. 
It is found that as the value of $c$ decreases from 137.03 to 134.28 a.u., the strength of SOC per Hf atom in HfTe increases from 383.71 to 1296.12 meV. Despite this change in SOC strength, the resulting topological phase remains exactly the same as that for $c$=137.03 (for details, see Table IV in the SM~\cite{spl}). 
Since band crossings above the Fermi level are typically not suitable for experimental observation using angle-resolved photoemission spectroscopy (ARPES), chemical doping is required to lower their energy and render them experimentally accessible. Therefore, we considered various doping combinations, including HfTe$_{0.875}$Se$_{0.125}$, HfTe$_{0.75}$Se$_{0.25}$, HfTe$_{0.625}$Se$_{0.375}$, and HfTe$_{0.5}$Se$_{0.5}$.
For the corresponding band structures of HfTe$_{1-x}$Se$_x$ ($x = 0.125$, $0.250$, $0.375$, and $0.5$), refer to Fig. 5 in the SM~\cite{spl}.
We have found that the doped materials exhibit the same topological excitations as in the parent HfTe material, which clearly demonstrates that the topological phases are robust against chemical doping (see Tables IV and V in the SM~\cite{spl} for details).
For the doped material HfTe$_{0.5}$Se$_{0.5}$, we found that the energy of TDF2 is lower than that of the parent compound HfTe. Similarly, for WP2, the energy is also reduced compared to the parent HfTe. This clearly suggests that doping can render these states experimentally accessible.
It is important to highlight that, in the present case, the nodal lines in HfSe and HfTe are symmetry enforced and originate from the underlying crystalline symmetries of the system. Substitutional doping between Se and Te preserves the same crystal symmetry and space group, as these elements occupy equivalent lattice sites. Consequently, any intermediate composition between Se and Te maintains the symmetry conditions required to protect the nodal lines. As a result, doping does not disrupt or gap out the nodal line features but mainly leads to a shift in the energy positions of the nodal features. Our calculated results clearly demonstrate this behavior.
Along with this, we have also explored the topological phases of ZrTe and TiTe materials in the absence and presence of SOC (the corresponding band structure is shown in Fig. 6 in the SM~\cite{spl}). In the absence of SOC, ZrTe and TiTe host TDF and nodal line fermions. Upon inclusion of SOC, both materials exhibit 24 Weyl fermions, two TDFs, and a straight nodal line (see Tables VII and VIII in the SM~\cite{spl}). Finally, we remark that the topological excitations present in the HfTe material are robust against various parameters and perturbations. Since the ZrTe material has already been synthesized, HfTe could also be prepared and is expected to exhibit the aforementioned topological excitations in ARPES experiments.

\begin{figure}
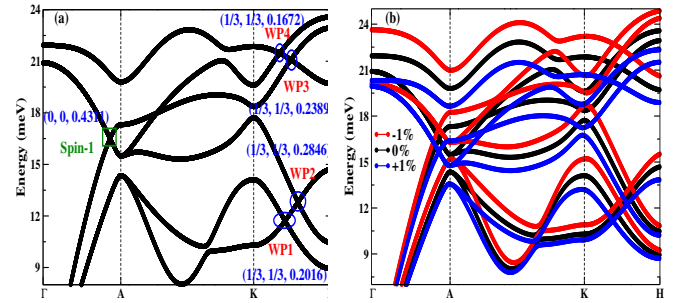

\centering
\includegraphics[width=0.49\linewidth, height=3.9cm]{Figure_eps/full_ph.eps}
\includegraphics[width=0.49\linewidth, height=3.85cm]{Figure_eps/compare_ph_band_main.eps}
\caption{\label{Fig.ph_band}\footnotesize{(Color online) The phonon dispersion of HfTe: (a) at optimized latice parameter (b) comparison of phonon modes with -1\%, 0\% and 1\% changes in the lattice parameter, along high-symmetric lines. WP represents the phonon Weyl point.}}
\end{figure}

{\it Topological phonons.|}
Beyond electrons, phonons, which are quantized vibrational states of atoms, play an important role in superconductivity, thermal conductivity, and thermoelectricity. Solids with $n$ atoms in their primitive unit cell have 3 acoustic and 3$n$-3 optical branches in their phonon dispersion, as calculated for HfTe in Fig. \ref{Fig.ph_band}. In a recent study~\cite{pandeyBCC}, it was observed that phonon bands are sensitive to supercell size~\cite{DFT_book_Pandey2026}. Therefore, we performed phonon calculations using 2$\times$2$\times$2 and 3$\times$3$\times$3 supercell sizes to examine this sensitivity. We found a negligible difference in the phonon bands between the 2$\times$2$\times$2 and 3$\times$3$\times$3 supercell sizes, indicating that the phonon bands are converged and sufficient to capture the phononic properties (see Fig. 7 in the SM~\cite{spl} for details). It is observed that only the phonon bands along the $\Gamma$-A direction are triply degenerate (at 16.50 meV), while the phonon bands along other high-symmetric directions in the Brillouin zone are doubly degenerate (see Fig. \ref{Fig.ph_band}(a)).
It is also important to mention that the presence or absence of phonon degenerate points along high-symmetric points, lines, and surfaces does not guarantee that such degenerate points will not be present at general points in the Brillouin zone. However, most studies are limited to high-symmetric paths, which limits the ability to predict new topological excitations. In this direction, searching for phonon degenerate points associated with two or more bands in the Brillouin zone is highly desirable, and the \texttt{PH-NODE}~\cite{PANDEY2024109281} code is well-equipped for this purpose. Using the \texttt{PH-NODE} code, we found Weyl phonons, spin-1 Weyl optical phonons, and nodal-line phonons (see Fig. \ref{Fig.ph_type}).
It is noted that, at present, there is no widely available \textit{first-principles}-based code that directly computes topological invariants, such as the Chern number, for phonon bands. Consequently, an explicit calculation of phonon topological invariants at the DFT level is not performed in the present work. Instead, we analyze the 3D phonon energy dispersion in the vicinity of the identified nodal points across relevant \textbf{k}-space regions. The characteristic features of these dispersions, such as linear band crossings, allow us to infer the expected topological nature of the phonon nodes. Our conclusions regarding the associated Chern numbers are therefore based on these dispersion features and are consistent with previously reported results obtained from effective tight-binding models in the literature~\cite{PhysRevB.97.054305}.

In order to check the nature of each phononic topological excitation, we have calculated the 3D energy dispersion around each one. We have found Weyl phonons of type-I, type-II, and nearly type-III, as well as type-II nodal line/arc phonons and spin-1 Weyl phonons (see Fig. \ref{Fig.ph_type} in the main text and Figs. 8-13 \& Table IX in the SM~\cite{spl}). 
In order to examine the robustness of the obtained topological excitation against variations in the lattice parameters, we varied the optimized lattice parameter by $\pm$1\% and and found no imaginary modes in the phonon dispersion plot. This indicates that the material remains mechanically stable under a $\pm$1\% change in lattice parameter. We found the same topological excitation under the change in lattice parameter as in the optimized lattice, which clearly demonstrates that the topological phase is robust against such a change in lattice parameter. In addition to this, we also performed phonon calculations for ZrTe and TiTe and found the aforementioned topological excitations in TiTe as well. However, in ZrTe, we did not find spin-1 Weyl phonons, although other topological phases are present. Interestingly, in TiTe, the spin-1 Weyl phonon is present between the optical branches, whereas in HfTe, it is found between the highest acoustic branch and the lower two optical branches (see Fig. 14 in the SM~\cite{spl} for details). The presence of such three-fold spin-1 phonon modes (with two optical and one acoustic) leads to a strong thermoelectric response, and these degeneracies significantly suppress the lattice thermal conductivity of the material~\cite{PhysRevMaterials.2.114204}. Moreover, the presence of three-fold fermionic excitations in the material simultaneously enhances the thermopower, and the combination of such electronic and phononic excitations is key to achieving high thermoelectric performance.
It is also important to understand the physical origin of the threefold spin-1 phonon nodal point observed near the A point in HfTe. We have found that this nodal point arises from atomic inversion. Bands 3, 4, and 5 are primarily contributed by Hf and Te atoms. Near the A point, the contribution from Te atoms is more than that from Hf atoms, which is completely opposite to the situation away from the A point. This atomic inversion is the key factor leading to the formation of the spin-1 nodal point in the phonon spectrum.



\begin{figure}\label{Fig.ph_type} 
\centering
\includegraphics[width=0.32\linewidth, height=3.5cm]{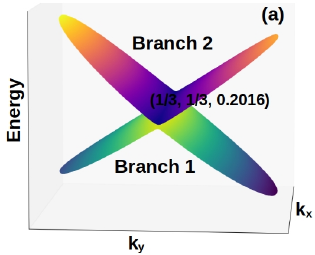}
\includegraphics[width=0.32\linewidth, height=3.5cm]{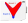}
\includegraphics[width=0.32\linewidth, height=3.5cm]{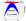}
\includegraphics[width=0.32\linewidth, height=3.5cm]{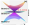}
\includegraphics[width=0.32\linewidth, height=3.5cm]{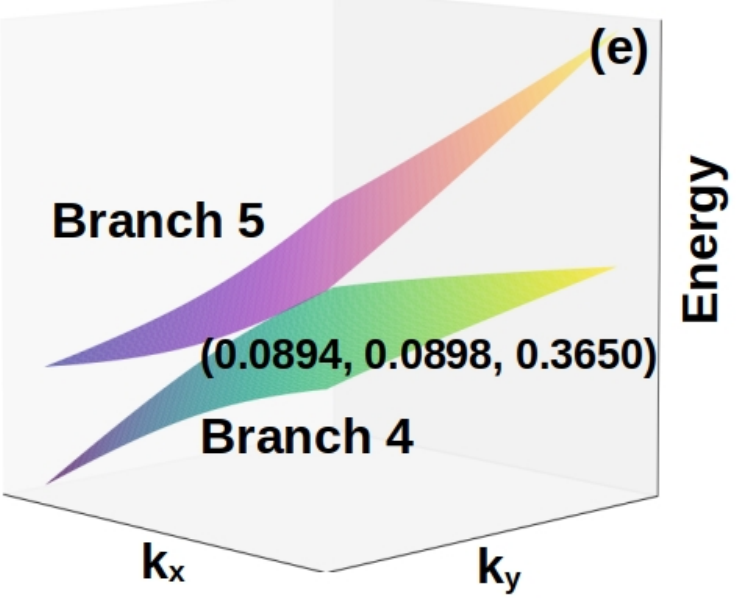}
\includegraphics[width=0.32\linewidth, height=3.5cm]{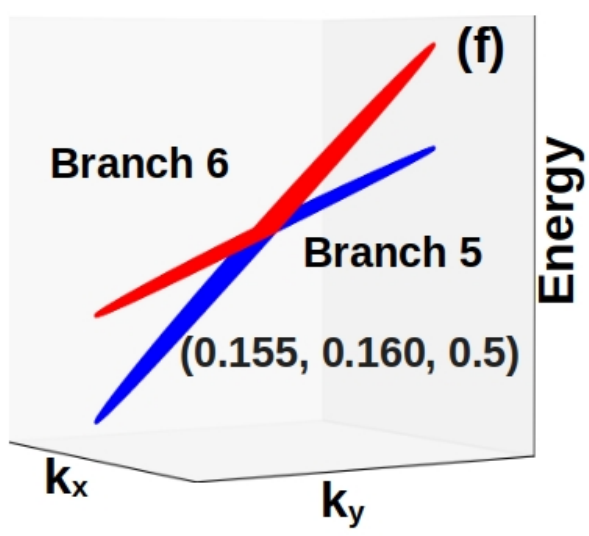}
\caption{\label{Fig.ph_type}\footnotesize{(Color online) 3D representation of the phonon energy dispersions of HfTe: (a) Type-I Weyl phonons (WP1, WP2, WP3, WP4) with topological charge, $C=\pm 1$; (b) Type-II Nodal line phonon; (c) Weyl phonon; (d) Three-branch spin-1 Weyl point with topological charges, $C=0, \pm 2$; (e) ((f)) Type-II Weyl phonons between branches 4-5 (5-6).}}
\end{figure}

{\it Conclusions.|}
In summary, we have performed a systematic theoretical investigation of electronic and phononic topological excitations in TTe (T=Hf, Zr \& Ti) using \textit{first-principles} DFT calculations. Our study unveils multiple unconventional topological excitations, including the presence of Weyl phase (with type-I, type-II, and type-III), nodal-line phase (with type-I, type-II, and straight line), and threefold pseudospin/spin-1 phase, in both the electronic and phononic excitation spectra of TTe. In particular, threefold spin-1 nodal point and symmetry-enforced straight nodal line are found along $\Gamma$–A path in the electronic spectrum, both in the presence and absence of SOC. Interestingly, the threefold spin-1 nodal point phonons are also observed along the $\Gamma$–A high-symmetric direction in the phononic spectrum. 
From our findings, we expect that such topological excitations are not exclusive to materials belonging to space groups 198 and 223, and may be present in other materials as well.
Our calculated topological excitations and their robustness against variations in XC functionals, the strength of SOC, small changes in lattice parameters, and chemical doping further confirm that the multiple topological phases in the TTe crystals are likely to be observed in experiments. 
The simultaneous presence of multiple topological phases in both electronic and phononic states in TTe materials opens exciting opportunities to deepen the fundamental understanding of topological physics. 

\bibliography{MS}
\bibliographystyle{apsrev4-2}

\end{document}